\documentclass[%
 reprint,
 twocolumn,
 showpacs,
 amsmath,amssymb,
 aps,
 prl,
]{revtex4-1}
\usepackage[usenames,dvipsnames,svgnames,table]{xcolor}
\usepackage{graphicx}
\usepackage[colorlinks=true,
            linkcolor=blue,
            urlcolor=blue,
            citecolor=blue]{hyperref}
\usepackage[utf8]{inputenc}
\usepackage[english]{babel}

\addto\captionsenglish{}

\def\p{{\partial}}
\def\Re{\mathop{\text{Re}}}
\def\Im{\mathop{\text{Im}}}

\DeclareMathOperator{\csch}{csch}
\RequirePackage{color}

\begin{document}

\title{Quantized Laplacian growth, II: 1D hydrodynamics of the Loewner density}
\author{Oleg Alekseev}
\email{teknoanarchy@gmail.com}
\date{\today}
\affiliation{%
Chebyshev Laboratory, Department of Mathematics and
Mechanics, Saint-Petersburg State University, 14th Line, 29b, 199178
Saint-Petersburg, Russia
}%
\begin{abstract}
	A systematic analytic treatment of fluctuations in Laplacian growth is given. The growth process is regularized by a short-distance cutoff $\hbar$ preventing the cusps production in a finite time. This regularization mechanism generates tiny inevitable fluctuations on a microscale, so that the interface dynamics becomes chaotic. The time evolution of fluctuations can be described by the universal Dyson Brownian motion, which reduces to the complex viscous Burgers equation in the hydrodynamic approximation. Because of the intrinsic instability of the interface dynamics, tiny fluctuations of the interface on a microscale generate universal patterns with well developed fjords and fingers in a long time asymptotic.
\end{abstract}
\maketitle 

A pattern formation in highly unstable, dissipative, nonlinear growth processes still posses a great challenge in nonequilibrium statistical physics. Although the role of noise on a microscale is believed to be crucial in the pattern formation phenomena~\cite{StanleyBook}, this mechanism has been never implemented for a broad class of growth processes known as Laplacian growth~\cite{PelceBook,BensimonRMP}.

The Laplacian growth problem embraces a variety of diffusion driven growth processes typically observed in physical, chemical and biological systems. The best known examples are the viscous fingering in a Hele-Shaw cell, when a less viscous fluid is injected into a more viscous one in a narrow gap between two plates~\cite{PelceBook,BensimonRMP}, and diffusion-limited aggregation, which is realized by tiny Brownian particles with size $\hbar$ diffusing and sticking to the boundary of the cluster~\cite{WittenSanderPRL,ErzanRMP}. The relation between these processes is still puzzling~\cite{Levitov98,AlekseevPRESLG}. A naive limit of the vanishing particle size, $\hbar\to0$, in diffusion-limited aggregation leads to the ill-defined Hele-Shaw problem, when the initially smooth boundaries quickly develop cusplike singularities~\cite{BensimonRMP}. Thus, a role of the short-distance cutoff in Laplacian growth should be clarified.

The nonzero $\hbar$ significantly affects the interface dynamics by cutting off cusplike singularities at the interface, and generating inevitable noise on a microscale. Remarkably, the integrable structure of the idealized problem survives~\cite{TeodorescuNPB}, and the Lauglin's theory of the quantum Hall effect provides a clue to a statistical description of Laplacian growth~\cite{QLG1}.

In this paper we consider the pattern formation in the regularized Laplacian growth problem. In particular, we clarify the formation of complex patterns with the well developed fjords and fingers.

\textit{Statistical theory of Laplacian growth.} The short-distance regularization of the Laplacian growth problem implies that the change of the area of the domain is quantized and equals an integer multiple of the area quanta $\hbar$. The domain can be then considered as an incompressible aggregate of tiny particles with the size $\hbar$ obeying the Pauli exclusion principle. The statistical properties of the aggregate are universal and common with quantum chaotic systems. Namely, the distribution of fluctuations at the boundary of the aggregate is described by the Dyson circular $\beta$-ensemble~\cite{QLG1}~\footnote{The distribution~\eqref{P} at $\beta=2$ was obtained in~\cite{QLG1}. Below, we will assume that $\beta$ is arbitrary, and comment about the case $\beta=2$ separately.},
\begin{equation}\label{P}
	P(\theta_1,\dotsc,\theta_N)\propto\prod_{i<j}\left|2\sin\left(\frac{\theta_i-\theta_j}{2}\right)\right|^\beta.
\end{equation}
The density of eigenvalues $\rho(\theta)=q\sum_{i=1}^N \delta(\theta-\theta_i)$ is a smooth function of the angle $\theta$ in the large $N$ limit, where $q=\hbar/\delta t$ is a quanta of the growth rate $Q=Nq$, so that $\rho(\theta)$ is normalized as $(2\pi)^{-1}\int_0^{2\pi}\rho(\phi)d\phi=Q$.

\begin{figure}[t!]
\centering
\includegraphics[width=1\columnwidth]{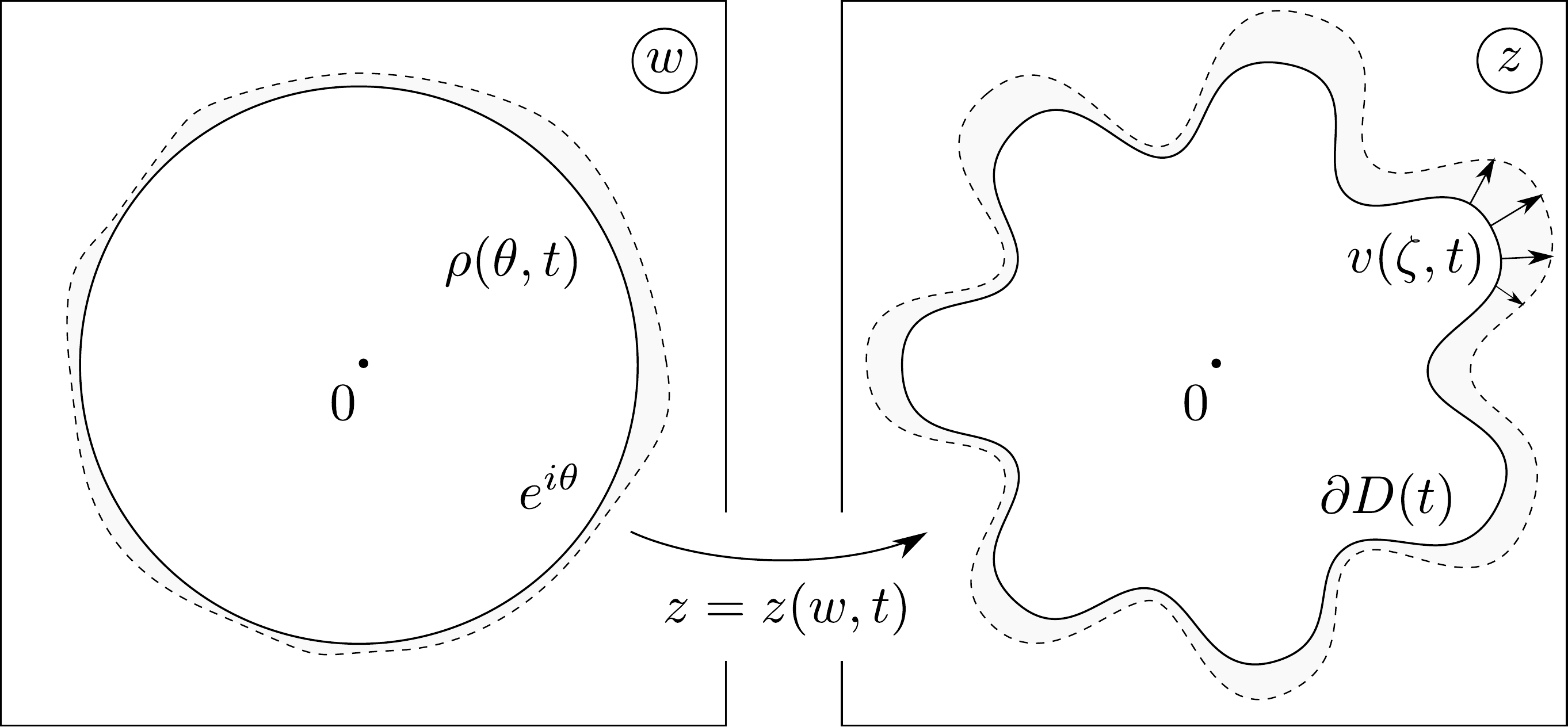}
\caption{\label{map2} The time dependent conformal map $z(w,t)$ from the complement of the unit disk in the auxiliary $w$ plane to the exterior of $D(t)$ in the physical $z$ plane. The dashed line in the $z$ plane represents the advance of the interface, $\Gamma(t)\to \Gamma(t+\delta t)$, per time unit $\delta t$. The distribution of normal velocities $v_n(\zeta,t)=|z'(e^{i \theta},t)|^{-1}\rho(\theta)$ along $\Gamma (t)$ is determined by the density of eigenvalues $\rho(\theta)$ of the Dyson ensemble.}
\end{figure}

Let $z(w,t)$ be the time dependent conformal map from the complement of the unit disk in the auxiliary $w$ plane to the exterior of the domain $D(t)$ in the physical $z$ plane, so that $z(\infty,t)=\infty$, and the conformal radius $r(t)=z'(\infty,t)$ is a positive function of time~\footnote{The dot and prime denote the partial derivatives with respect to time and the coordinate correspondingly.}. The advance of the boundary $\Gamma(t)=\p D(t)$ in the $z$ plane, $\delta h(\zeta,t)=v_n(\zeta,t)\delta t$, where $v_n(\zeta,t)$ is the normal interface velocity, is related to the density of eigenvalues $\rho(\theta)$ by the conformal factor~\cite{QLG1} (see Fig.~\ref{map2}),
\begin{equation}\label{v}
	v_n(\zeta,t)=|z'(e^{i \theta},t)|^{-1}\rho(\theta),\quad \zeta=z(e^{i \theta},t).
\end{equation}

Eq.~\eqref{v} can be recast in Loewner-Kufarev equation, which describes a sequence of conformal maps of subordinal domains parametrized by time $t$~\cite{Loewner23,Kufarev}, 
\begin{equation}\label{LK-eq}
	\frac{\p z(w,t)}{\p t}=wz'(w,t)\int_0^{2\pi}\frac{d\phi}{2\pi}\frac{w+e^{i \phi}}{w-e^{i\phi}}\frac{\rho(\phi)}{|z'(e^{i\phi},t)|^2}.
\end{equation}
Since $\rho(\phi)=q\sum_{i=1}^N \delta(\phi-\theta_i)$, the positions of eigenvalues $\theta_i$ represent the points which provide growth. In particular, the case when $\rho\propto|z'(e^{i \theta_0},t)|^2\delta(\phi-\theta_0)$ is the normalized Dirac peak corresponds to the \textit{local} Loewner evolution~\cite{Loewner23}, while the uniform distribution of eigenvalues, $\rho(\phi)=Q$, generates the \textit{non-local} deterministic Laplacian growth with the growth rate $Q$. 

\textit{Evolution of the Loewner density.} We introduce the time evolution of the Loewner density, $\rho(\phi)\to\rho(\phi,t)$ in eq.~\eqref{LK-eq}, by using the following two interpretations of the distribution~\eqref{P}: (1) $P$ is the ground state wave function for the \textit{conservative} Calogero  model defined on the unit circle by the Hamiltonian
\begin{equation}\label{CS-H}
	\mathcal H=-\sum_{j=1}^N\frac{\p^2}{\p \theta_j^2}+\frac{1}{4}\sum_{k\neq j}^N\frac{\beta(\beta/2-1)}{\sin^2(\theta_k-\theta_j)/2},
\end{equation}
and (2) $P$ is the probability distribution function of $N$ charged Brownian particles, subjected to an electric force $E(\theta_i)=-\p_{\theta_i} W$ with $W=-\sum_{i<j}\log|\sin(\theta_i-\theta_j)/2|$, and friction with strength $\gamma$ (we will set $\gamma=1$ below). At temperature $\beta^{-1}$ during a small ``time'' interval $\delta \tau$ changes in particle positions $\delta \theta_i$ are given by
\begin{equation}\label{d-theta}
	\langle \delta \theta_i\rangle= q E(\theta_i)\delta \tau,\qquad \langle(\delta \theta_i)^2\rangle=2 q \beta \delta \tau,
\end{equation}
and all higher moments are zero. The  joint probability density $P(\theta_1,\dotsc,\theta_N;\tau)$ then satisfies the Fokker-Planck equation~\cite{DysonJMPBR},
\begin{equation}\label{FP}
	q^{-1}\frac{\p P}{\p \tau}=\mathcal L P,\qquad \mathcal L=\sum_i\frac{\p}{\p \theta_i}\left(\beta^{-1}\frac{\p}{\p \theta_i}+\frac{\p W}{\p \theta_i}\right),
\end{equation}
and its unique stationary solution is~\eqref{P}. Note, that the ``time'' $\tau$ is a parameter of the Dyson model. Its relation with the physical time $t$ will be clarified below. Although the transformation $\tilde P=P\exp(-W/2)$ recasts the Fokker-Plank operator~\eqref{FP} in the Calogero Hamiltonian~\eqref{CS-H}, the system~\eqref{FP} is \textit{dissipative}.

\textit{Hydrodynamical description.} In the large $N$ limit exact collective descriptions of both models,~\eqref{CS-H} and~\eqref{FP}, are known and have the hydrodynamical form closely related to the Hopf equation,
\begin{equation}\label{Hopf}
	\p_{\tau}u+u\p_\theta u=0,
\end{equation}
where $u=v+i\pi \rho$ (outside the unit disk) is a sum of the density $\rho$ and velocity $v$ operators in the Calogero model~\eqref{CS-H}, while in the Dyson model $u=\rho^H-i\rho$ is the Cauchy transform of the density. By $\rho^H(\theta)=P\int_0^{2\pi}\rho(\theta')\cot[(\theta-\theta')/2]d \theta'$ we denoted the Hilbert transform of $\rho(\theta)$.

The nonlinearity of the Hopf equation results in the formation of shock waves, which must be regularized either by dispersion (in the Calogero model), or by dissipation (in the Dyson model). In the former case one obtains the \textit{Benjamin-Ono} equation as an effective description of Calogero hydrodynamics in the limit of weak nonlinearity and dispersion~\cite{AW05,ABW09},
\begin{equation}\label{BO}
	\p_\tau u +u\p_\theta u=-\alpha_{BO}\p^2_\theta u^H,
\end{equation}
where $2\alpha_{BO}=\sqrt {\beta/2}-\sqrt {2/\beta}$. In the latter case one arrives at the \textit{complex Burgers} equation,
\begin{equation}\label{BE}
	\p_\tau u+u\p_{\theta}u=\nu \p^2_\theta u,
\end{equation}
with $\nu=(1-\beta/2)q$, which describes the hydrodynamic limit of the Dyson model as the interplay between nonlinearity and dissipation. Although the Benjamin-Ono term, $\alpha_{BO}\p_\theta^2 u^H$ is of the same order as the Burgers term, $\nu\p_\theta^2 u$, it is very different in nature. It results in the real correction, $\delta\omega\sim \alpha_{BO}k|k|$, to the spectrum of linear waves, while the Burgers term leads to the dissipation, $\delta\omega\sim-i \nu k^2$.

\textit{Semiclassical limit.} At $\beta=2$ both coefficients, $\alpha_{BO}$ and $\nu$, vanish. However, the semiclassical limits of quantum hydrodynamical equations are subtle. The semiclassical limit of eq.~\eqref{BO} requires a shifting $2\alpha_0\to\sqrt{\beta/2}$. Thus, $\alpha_{BO}$ does not vanish for free fermions~\cite{AW05,Jev92}. We assume that the shifting of parameters is also required in eq.~\eqref{BE}~\footnote{We will address this issues elsewhere.}. Therefore, we consider $\nu$ ($0<\nu\ll1$) as the ``effective'' short-distance cutoff, so that the term, $\nu \p^2_\theta u$, survives at $\beta=2$ and flattens the shock waves.

Solutions to both equations,~\eqref{BO} and~\eqref{BE}, can be coupled to the Loewner-Kufarev equation~\eqref{LK-eq} thus generating a non-trivial evolution of the domain. In this paper we will  only consider the Dyson model, because the Laplacian growth is a purely dissipative process~\footnote{Solutions to the Benjamin-Ono equation are relevant for the evolution of conservative systems, e.g., electronic droplets in the quantum Hall regime. Below, we will introduce a general procedure, which can be used to study these systems as well.}. The dissipative term in eq.~\eqref{BE}, which has a statistical origin, can be connected to the oil viscosity, $\mu_\text{oil}$, as $\nu\sim\mu^{-1}_\text{oil}$.

\textit{Dissipation of fluctuations}.  In terms of the density, $\rho=-\Im u$, the complex Burgers equation~\eqref{BE} takes the following form:
\begin{equation}\label{rho-eq}
	\p_\tau\rho+\p_\theta(\rho \rho^H)=\nu \p_\theta^2\rho.
\end{equation}
The exact analytical formulas for solutions to the transport equation~\eqref{rho-eq} can be obtained by the method of pole expansion~\cite{ChoodnovskyNCB,MatsunoJMP}. The idea is to consider the solutions $\rho(\theta,\tau)$ as the functions in the complex plane (we only consider the meromorphic solutions),
\begin{equation}\label{rho}
	\rho(\theta,t)=Q-2\nu\Re\sum_{k=1}^M\frac{\xi_k(t)}{e^{i \theta}-\xi_k(t)},
\end{equation}
where the poles at $\xi_k$ lie inside the unit disk in the $w$ plane, and the normalization, $(2\pi)^{-1}\int_0^{2\pi}\rho(\phi,t)d\phi=Q$, is assumed. Besides, we introduced the physical time $t=t(\tau)$, so that in the \textit{long time asymptotic} one has,
\begin{equation}\label{time}
	d \tau(t)/d t=2(\nu/q)\exp\left[-2r(t)/\delta\right],
\end{equation}
where $r(t)$ and $\delta$ are the conformal radius and characteristic scale (e.g., the typical width of fjords) of the domain correspondingly. The proposition~\eqref{time} can be justified from the statistical mechanic point of view, namely, by coupling stochastic Laplacian growth to conformal field theories and finding the conditions for certain objects to be martingales~\cite{QLG3}.

The anzats~\eqref{rho} is a solution of the transport equation~\eqref{rho-eq}, if and only if the following system of coupled first order differential equations is satisfied:
\begin{equation}\label{xi}
	\frac{d\phi_k}{d \tau}=-Q+\nu\sum^M_{l\neq k}\coth\frac{\phi_k-\phi_l}{2}, \quad \phi_k=\log \xi_k,
\end{equation}
Thus, the non-linear partial differential equation~\eqref{rho-eq} reduces to the many body problem of $M$ interacting particles with the coordinates $\xi_k$ inside the unit disk. The particles attract each other in the tangential direction, and repel each other in the radial direction. They tend to form radial lines, eventually coalescing into a single point at the origin (because of the constant drift $-Q$) with a characteristic ``lifetime'' $\tau^*\sim Q^{-1}$. Thus, any initial density relaxes to the uniform distribution.

The dynamical system~\eqref{xi} can be embedded into the Calogero $M$ body problem. By taking  the time derivative of eq.~\eqref{xi} one obtains
\begin{equation}
	\frac{d^2}{d \tau^2}\phi_k=-\nu^2\sum_{l\neq k}\coth\frac{\phi_k-\phi_l}{2}\csch^2\frac{\phi_k-\phi_l}{2}.
\end{equation}
Note, however, that the phase volume of the system~\eqref{xi} shrinks with time, because of the dissipative structure of the complex Burgers equation.

The poles $\xi_k$ are the collective coordinates parameterizing  fluctuations of the interface. Because of the short-distance cutoff $\hbar$, the surface motion is jumpy and grainy. The methods of statistical mechanics allows one to reduce microscopic degrees of freedom (fluctuations of the interface) to a set of collective coordinates, while the remaining degrees of freedom transform into effective noise. The hydrodynamic equation~\eqref{rho-eq} implies an averaging with respect to noise, so that only the deterministic dynamics of the collective coordinates survives~\eqref{xi}. This is a common feature of the mean field approach often applied to study the behavior of complex stochastic systems.

\textit{Coupling the Burgers hydrodynamics with the Loewner-Kufarev equation.} The solutions $\rho(\theta,t)$ to the transport equation~\eqref{rho-eq} determine the interface dynamics in Laplacian growth~\eqref{v}. Rewriting the normal interface velocity in terms of the conformal map, $v_n(e^{i \theta},t)=\Im(\overline{\p_t z} \p_\theta z)/|\p_\theta z|$, and using eqs.~\eqref{v},~\eqref{rho}, one obtains the following equation of motion of the boundary,
\begin{equation}\label{slg}
	\Im\left[\overline {\p_tz(e^{i \theta},t)}\p_\theta z(e^{i \theta},t)\right]=Q-2\nu\Re\sum_{k=1}^M\frac{\xi_k}{e^{i \theta}-\xi_k},
\end{equation}
This equation has a straightforward interpretation. In the case when $z_k(t)\equiv z(1/\bar \xi_k(t),t)=const$, eq.~\eqref{slg} describes deterministic Laplacian growth with $1+M$ oil sources with the rates $Q$ and $-\nu$ located at the points $z_{\infty}=\infty$ and $z_k$'s. When the dynamics of the poles $\xi_k$ is determined by eq.~\eqref{xi}, their images, $z_k(t)$, in the $z$ plane move with time. Therefore, the evolution of the fluctuating boundary~\eqref{slg} has the form of the Laplacian growth equation with dynamical \textit{virtual} sources~\footnote{The pressure field in oil is regular at the points $z_k$'s. The only source providing the growth is located at infinity.}. 

\textit{Evolution of the Schwarz function.} The interface dynamics can be studied by using the Schwarz function approach~\cite{HowisonEJAM}. The Schwarz function $\mathcal S(z,t)$ for a sufficiently smooth curve $\Gamma(t)$ drawn on the complex plane is an analytic function in a striplike neighborhood the curve, such that $\bar z=\mathcal S(z,t)$ for $z\in \Gamma(t)$~\cite{DavisBook}. If the curve evolves in time, so does its Schwarz function. The Laplacian growth equation~\eqref{slg} can be rewritten in terms of the Schwarz function as follows:
\begin{equation}\label{SW}
	\p_t\mathcal S(z,t)=2\p_z\mathcal W(z,t),
\end{equation}
where the complex potential $\mathcal W=-p+i\psi$ is a sum of the negative pressure $p$ and the stream function $\psi$. The function $\mathcal S=\mathcal S^++\mathcal S^-$ (and $\mathcal W$ correspondingly) can be decomposed into a sum of two functions, $S^+$ and $\mathcal S^-$, which are regular in $D(t)$ and $\mathbb C\setminus D(t)$ respectively.

In the deterministic Laplacian growth $\p_z\mathcal W$ has no singularities in the oil domain, so that $\dot {\mathcal S}^+(z,t)=0$. If the boundary fluctuates with time, then eq.~\eqref{slg} implies $\dot {\mathcal S}^+(z,t)=2\nu\sum_k\log(z-z_k)$. Therefore, the dynamics of virtual sources located at $z_k(t)\in\mathbb C\setminus D(t)$ generates the branch cuts $\Gamma_k$ of the Schwarz function in the oil domain, which can be written as a sum of Cauchy type integrals,
\begin{equation}\label{S+}
	\mathcal S^+(z,t)=\mathcal S^+(z,0)+2\nu\sum_k\int_{z_k(0)}^{z_k(t)}\frac{\mathcal P_k(l)d l}{z-l},
\end{equation}
with Cauchy densities $\mathcal P_k(l)$  along the cuts~\cite{AMZ09}. The representation~\eqref{S+} implies that $\mathcal S^+$ has discontinuities across the cuts, $[\mathcal S^+(z)]_\text{cut}=\mathcal P_k(z)$, $z\in \Gamma_k$. The Cauchy densities are determined by the absolute velocities of the virtual sources $v(t)=dl(t)/dt$ in the oil domain, namely, $P_k(l)=v_k^{-1}(t)$. In particular, the Schwarz function has the logarithmic branch cut $\Gamma_k$, provided that $v_k(t)=const$.

\begin{figure}[t!]
\centering
\includegraphics[width=1\columnwidth]{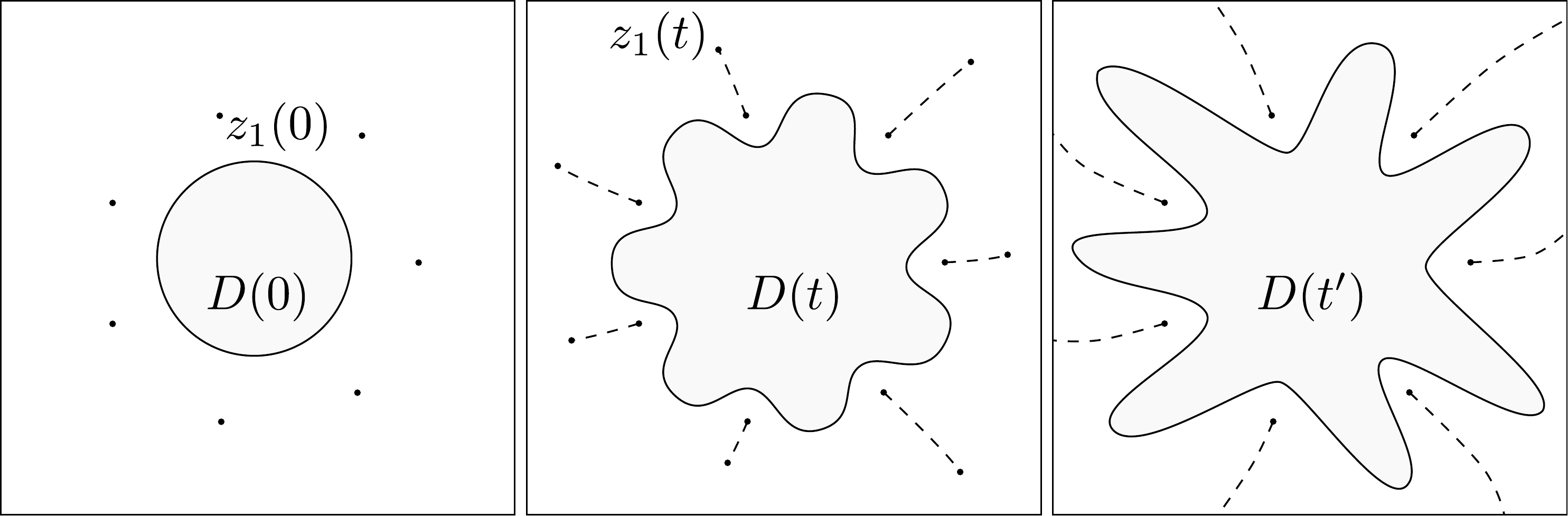}
\caption{\label{fjords}The dynamical generation of the fjords of oil, which separate fingers of the water droplet, is schematically depicted in three successive instants of time $0<t<t'$. The centerlines of the \textit{dynamically} generated fjords (the dashed curves outside the growing domain) are the branch cuts of $\mathcal S^+(z,t)$.}
\end{figure}

\textit{Formation of fjords.} Although the fluctuations of the interface velocity are small, $\nu\ll Q$ in eq.~\eqref{slg}, they considerably effect the highly unstable growth process. Namely, they result in the formation of deep fjords of oil separating the fingers of water (see Fig.~\ref{fjords}), so that the centerlines of the fjords~\footnote{The centerline is the imaginary line, which is equidistant from the fjord edges.} are the branch cuts of the Schwarz function~\eqref{S+}. Let us briefly mention some qualitative features of the fjord's evolution.

The Cauchy densities vary along the branch cuts thus resulting in the development of fjords with curved walls and nonzero opening angles. The characteristic width of the $k$th fjord is $\delta_k\simeq\nu/v_k$. The dynamics of poles~\eqref{xi} shows a tendency of small fjords to coalesce and form the large ones. Besides, in regularized Laplacian growth,~\eqref{slg} and~\eqref{xi}, the harmonic moments of the domain,
\begin{equation}
	t_n(t)=\frac{1}{2\pi n}\int_{\mathbb C \setminus D(t)}z^{-n}d^2z,\qquad n=1,2,\dotsc,
\end{equation}
decay with time, $\dot t_n=-(\nu/n) \sum_k z_k^{-n}(t)$, in contrast to the idealized problem when $\dot t_n=0$. The  experimentally observed decay of $t_n$'s was previously related with the effect of surface tension~\cite{SwinneyPRE}. Here, we show that the short-distance regularization of the idealized problem implies a similar effect.

\textit{Conclusion.} The short-distance regularization of the idealized Laplacian growth problem was considered and shown to possesses the following features:

(1) \textit{The interface remains smooth during the whole evolution.} For a large class of fluctuations of the initially smooth regular boundary, the solutions to the system of equations~\eqref{xi} and~\eqref{slg} remain regular for all times. Thus, the cusps formation is prevented.

(2) \textit{The complex irregular shapes are observed at large times.} The growth process~\eqref{slg} is highly sensitive to tiny perturbations of the boundary, which yield the tip splitting and the development of fjords. The fluctuations constantly appear during the evolution, thus generating an extremely complex structure of viscous fingers, when the tips repeatedly split and form new fingers and fjords.

(3) The chaotic dynamics of the interface in Laplacian growth is described statistically as a relaxation to equilibrium in the Dyson $\beta$-ensemble.


\bibliography{biblio}{}
\end{document}